# Towards a Unified Framework for Determining Conformational Ensembles of Disordered Proteins

Hamidreza Ghafouri[1], Pavel Kadeřávek[2,3], Ana M Melo[4,5], Maria Cristina Aspromonte[1], Pau Bernadó[6], Juan Cortés[7], Zsuzsanna Dosztányi[8], Gabor Erdos[8], Michael Feig[9], Giacomo Janson[9], Kresten Lindorff-Larsen[10], Frans A. A. Mulder[11], Peter Nagy[12,13,14], Richard Pestell[15,16,17,18,14,19], Damiano Piovesan[1], Marco Schiavina[20], Benjamin Schuler[21,22], Nathalie Sibille[23], Giulio Tesei[10], Peter Tompa[18,24], Michele Vendruscolo[25], Jiri Vondrasek[26], Wim Vranken[27,28,29,30,31], Lukas Zidek[2,3], Silvio C.E. Tosatto[1,*], Alexander Miguel Monzon[1*]

*Corresponding authors: Department of Biomedical Sciences, University of Padova, Padova 35121, Italy. E-mails: alexander.monzon@unipd.it and silvio.tosatto@unipd.it.

[1] Department of Biomedical Sciences, University of Padova, Italy. Via Ugo Bassi 58/B, 35131 Padova, Italy

[2] Central European Institute of Technology (CEITEC), Masaryk University, Kamenice 5, 62500 Brno, Czech Republic

[3] National Centre for Biomolecular Research, Faculty of Science, Masaryk University, Kamenice 5, 62500 Brno, Czech Republic

[4] iBB—Institute for Bioengineering and Biosciences, Instituto Superior Técnico, Universidade de Lisboa, Av. Rovisco Pais, 1049-001 Lisboa, Portugal

[5] Associate Laboratory i4HB—Institute for Health and Bioeconomy at Instituto Superior Técnico, Universidade de Lisboa, Av. Rovisco Pais, 1049-001 Lisboa, Portugal

[6] Centre de Biologie Structurale (CBS), Université de Montpellier, INSERM and CNRS, 34090 Montpellier, France.

[7] LAAS-CNRS, Université de Toulouse, CNRS, F-31400 Toulouse, France

[8] Department of Biochemistry, ELTE Eötvös Loránd University, Budapest

[9] Department of Biochemistry and Molecular Biology, Michigan State University, East Lansing, Michigan 48824, United States

[10] Structural Biology and NMR Laboratory, Linderstrøm-Lang Centre for Protein Science, Department of Biology, University of Copenhagen, Ole Maaløes Vej 5, DK-2200 Copenhagen, Denmark

[11] Johannes Kepler University, Institute of Biochemistry, Linz 4040, Austria

[12] Department of Molecular Immunology and Toxicology and the National Tumor Biology Laboratory, National Institute of Oncology 1122 Budapest, Hungary

[13] Department of Anatomy and Histology, HUN-REN–UVMB Laboratory of Redox Biology Research Group, University of Veterinary Medicine, 1078 Budapest, Hungary

[14] Chemistry Coordinating Institute, University of Debrecen, 4012 Debrecen, Hungary

[15] Baruch S. Blumberg Institute, Doylestown, PA, 18902, USA

[16] Xavier University School of Medicine at Aruba, Oranjestad, Aruba
[17] The Wistar Institute, Philadelphia, PA, 19107, USA
[18] HUN-REN Office for Supported Research Groups (TKI), Cell Cycle Laboratory, National Institute of Oncology, 1122 Budapest, Hungary
[19] Semmelweis University, Budapest, 1117 Hungary
[20] Magnetic Resonance Center and Department of Chemistry “Ugo Schiff”, University of Florence, Via Luigi Sacconi 6, 50019 Sesto Fiorentino, Florence, Italy
[21] Department of Biochemistry, University of Zurich, Zurich, Switzerland
[22] Department of Physics, University of Zurich, Zurich, Switzerland
[23] Centre de Biologie Structurale (CBS), CNRS - Univ. Montpellier - Inserm, 34090 Montpellier, France
[24] Institute of Molecular Life Sciences, HUN-REN Research Centre for Natural Sciences, Budapest 1117, Hungary
[25] Yusuf Hamied Department of Chemistry, Centre for Misfolding Diseases, University of Cambridge, Cambridge, UK
[26] Institute of Organic Chemistry and Biochemistry, Czech Academy of Sciences, Flemingovo namesti 2, Prague 6, Czech republic
[27] Structural Biology Brussels, Bio-engineering Department, Vrije Universiteit Brussel, Elsene 1050, Belgium
[28] Interuniversity Institute of Bioinformatics in Brussels, Vrije Universiteit Brussel, 1050 Brussels, Belgium
[29] AI Lab, Vrije Universiteit Brussel, 1050 Brussels, Belgium
[30] Chemistry department, Vrije Universiteit Brussel, 1050 Brussels, Belgium
[31] Biomedical sciences, Vrije Universiteit Brussel, 1050 Brussels, Belgium

# Abstract

Disordered proteins play essential roles in myriad cellular processes, yet their structural characterization remains a major challenge due to their dynamic and heterogeneous nature. Here we present a community-driven initiative to address this problem by advocating a unified framework for determining conformational ensembles of disordered proteins. Our aim is to integrate state-of-the-art experimental techniques with advanced computational methods, including knowledge-based sampling, enhanced molecular dynamics, and machine learning models. The modular framework comprises three interconnected components: experimental data acquisition, computational ensemble generation, and validation. The systematic development of this framework will ensure the accurate and reproducible determination of conformational ensembles of disordered proteins. We highlight the open challenges necessary to achieve this goal, including force field accuracy, efficient sampling, and environmental dependence, advocating for collaborative benchmarking and standardized protocols.

# Introduction

Intrinsically disordered proteins and proteins including intrinsically disordered regions (IDPs) are prevalent in eukaryotic proteomes, comprising an estimated one third of the human proteome[1]. These proteins exist as conformational ensembles of rapidly interconverting conformations[2]. IDPs are implicated in a broad range of signaling and regulatory functions, such as signal transduction, gene expression, genome organization, cell-cycle control and the formation of biomolecular condensates[3]. Consequently, IDPs are central targets for biological and medical research.

A defining feature of IDPs is the link between the features of their conformational ensembles and their biological functions[4–6]. Understanding this relationship requires a robust framework for generating and analyzing conformational ensembles by combining computational and experimental techniques. In this perspective, we advocate a unified framework for generating, analyzing, and validating conformational ensembles of IDPs. This framework, developed through extensive discussions initiated during a workshop held in Prague in May 2023 (URL: https://ml4ngp.eu/workshop-prague), is structured around three key modules: (1) experimental observations informative of conformational ensembles, (2) computational generation of conformational ensembles consistent with the experimental information, and (3) validation and comparison of the resulting conformational ensembles. By integrating insights from different methodologies in structural and computational biology, our goal is to progress toward establishing standardized processes that support integrative studies of IDPs.

This modular framework (**Fig. 1**) underscores the importance of interdisciplinary collaboration in addressing the complexities of IDPs. By standardizing methodologies and establishing consensus within the scientific community, this framework seeks to enhance our understanding of the functions of IDPs, advance methodological rigor, and accelerate therapeutic discovery.

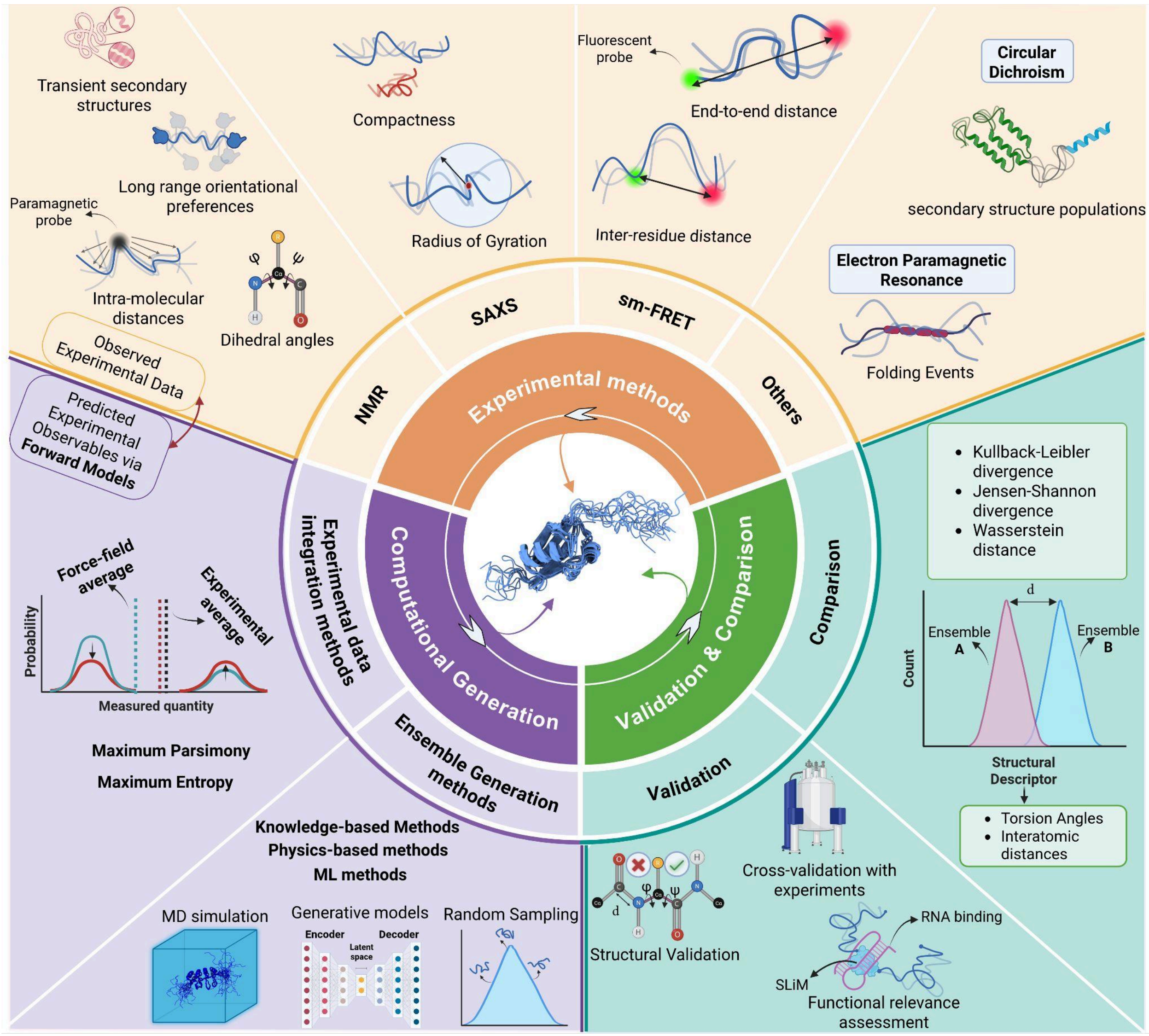

**Fig. 1. Schematic representation of the modular framework for the determination of conformational ensembles of IDPs.**

The illustration presents the modular framework for determining conformational ensembles of IDPs, comprising three interconnected modules: experimental methods, computational ensemble generation, and validation and comparison.The first module focuses on experimental techniques such as NMR spectroscopy, single-molecule FRET, SAXS, circular dichroism and electron paramagnetic resonance spectroscopy. These methods provide local and global structural information for characterizing the heterogeneous and dynamic nature of IDPs.The second module of the framework addresses the computational determination of conformational ensembles of IDPs. This module involves the generation of conformational ensembles and the integration of experimental data obtained in the first module. Ensemble generation methods can be categorized into physics-based (molecular dynamics simulations), knowledge-based, and machine-learning methods. In addition, maximum parsimony and maximum entropy are the main approaches for the integration of experimental data. The role of forward models is crucial in this step since they link the experimental observables to the generated models to ensure the goodness of fit to observed experimental data. Lastly, the validation and comparison tools complete the circular process of determining IDP conformational ensembles. Validation can be achieved through compliance of structures in the ensemble with basic physicochemical rules (structural validation), cross-validation using complementary experimental data, and functional relevance assessment. Comparison tools particularly involve statistical methods for analyzing distributions, such as Kullback-Leibler, Jensen-Shannon divergence scores and Wasserstein distances.

**Box 1. Glossary of terms used within the scope of this paper**

| Term | Definition |
|---|---|
| Intrinsically disordered proteins and disordered protein regions (IDPs) | Full-length proteins or protein regions that lack a stable ordered three-dimensional structure under physiological conditions |
| Conformers | Structures able to interconvert without making or breaking covalent bonds |
| Conformational collection | A set of conformers |
| Statistical ensemble | A conformational collection endowed with statistical weights. |
| Conformational state | A thermodynamic state of a protein described by a statistical ensemble |
| Forward Model | A computational framework that predicts experimental observables based on protein structures |

# Experimental techniques for studying IDP structural ensembles

## Nuclear magnetic resonance (NMR) spectroscopy

NMR spectroscopy is uniquely suited for studying IDPs in solution, offering atomic-level information. This technique can capture the dynamic and heterogeneous nature of IDPs by exploiting the sensitivity of nuclear magnetic moments to their local magnetic environment. This sensitivity enables precise, selective, and minimally invasive atomic-resolution assessment of structure, dynamics, and interactions of IDPs under near-physiological conditions[7], typically following isotope labeling. As the free energy landscape is shallow for IDPs, NMR observables are time-averages over the different conformations adopted by the IDPs due to the relatively slow (e.g. milliseconds) NMR detection. As a result, NMR measurements typically accurately reflect the conformational state, in the sense that observables are representative of the statistical ensemble. However, the lack of the information on individual states can be viewed as a fundamental limitation. The ensemble-averaged NMR parameters do not enable the specific reconstruction of individual conformers of the ensemble. Instead, the NMR parameters can be compared to those derived by forward models from simulated structural ensembles and thus benchmark the latter. Also, there are cases when some states of IDPs can be distinguished (e.g., the presence of cis peptide bonds, which give rise to sets of additional signals[8–10]), or when the NMR spectrum clearly carries a mark of their distribution (e.g., dynamics of molten globules manifested by line broadening[11]). NMR can also unveil sparsely populated NMR-invisible structural states by exploiting the transfer of magnetization between these states and NMR-visible states, through techniques like saturation transfer difference (STD) NMR and relaxation dispersion (RD) NMR. By analyzing the transferred magnetization, these techniques provide structural and dynamic information about minor states populated during structural transitions upon binding[12]. Moreover, NMR methods of selective detection of certain states have been designed (e.g. paramagnetic relaxation enhancement discussed below).

## Key NMR observables in IDP studies

Several NMR parameters can be directly correlated to conformational properties of disordered proteins. **Chemical shifts** reflect local electronic environments and provide insights into secondary structure propensities (SSP)[13–15]. Computational tools such as ShiftX[16], SPARTA+[17], PPM[18], and CamShift[19], include forward models to predict chemical shifts from conformational ensembles (Table 1).

**J-couplings**, particularly three-bond couplings, directly report on backbone torsion angles through the Karplus equation[20]. These measurements are limited by the accuracy of the corresponding forward models and the availability of measurable J couplings for specific torsion angles. For example, φ angles can reliably be probed[21]. Furthermore, ψ angles often require complementary approaches such as cross-correlated relaxation [22,23].

**Residual dipolar couplings (RDCs)** provide highly sensitive insights into IDP conformational preferences by reporting on the average orientation of bond vectors across all coexisting structures in solution[24,25]. RDCs reveal local secondary structure biases, having an opposite sign in alpha-helices and extended structures. However, their detailed structural interpretation requires calculating or fitting the alignment tensor, which is particularly challenging for IDPs. To avoid this calculation, forward models that use the laboratory reference frame have been proposed[26]. Various alignment media exist, yet careful selection is crucial to avoid unwanted interactions that could alter measurements. Another use of RDCs is for the cross-validation of conformational ensembles derived from other NMR and SAXS data, ensuring consistency in structural interpretations[27].

**Relaxation rates** offer insights into IDP dynamics across timescales. Through IDP-specific model-free formalism, the analysis of several backbone $^{15}N$ relaxation experiments in multiple magnetic fields probes the local and segmental motional amplitudes and their associated correlation times[28,29]. These data are also valuable for benchmarking molecular dynamics (MD) simulations, in which ensemble reweighting may be required to match experimental relaxation rates[30].

**Paramagnetic relaxation enhancement (PRE)** is particularly powerful for disordered proteins, revealing transient long-range contacts and compact conformations within heterogeneous ensembles. For instance, PRE measurements have revealed transiently folded states in different IDPs[31–33], complementing global techniques like SAXS in characterizing conformational distributions[34–36]. Paramagnetic relaxation interference (PRI) methods further expand this analysis, identifying correlated or anti-correlated motions in transiently compacted regions[37,38].

Since IDPs are highly sensitive to their environments, **in-cell NMR** is particularly indicated to study their context-dependent behavior under native-like conditions, such as in bacterial and eukaryotic cells. These studies, while technically challenging due to low sensitivity and high background noise, highlight the potential of NMR to bridge *in vitro* and *in vivo* IDP investigations[39–42].

# Single-molecule Förster resonance energy transfer (smFRET)

smFRET is a powerful tool for investigating the conformational dynamics and structural heterogeneity of IDPs. FRET, a non-radiative energy transfer from an excited donor fluorophore to an acceptor fluorophore, is highly sensitive to donor-acceptor distances, following an inverse sixth-power relationship. This makes smFRET a spectroscopic ruler capable of measuring inter- and intramolecular distances within a 2–10 nm range[43]. Unlike ensemble-averaging methods, smFRET provides a unique ability to resolve subpopulations, detect rare states, and capture dynamic transitions of individual molecules, making it particularly suited for IDP studies[44,45].

## smFRET applications to IDPs

smFRET has been instrumental in studying IDPs under diverse conditions. It has been employed to elucidate the conformational ensembles of IDPs in dilute solution[44,46,47] and crowded environments[48], including phase-separated biomolecular condensates[49,50] and live cells[51]. Furthermore, this technique has shed light on dynamic IDP complexes, revealing the interplay of transient structural states during functional interactions[45]. Single-molecule studies can be performed down to picomolar concentrations[43], avoiding the complications of aggregation and allowing functional/dysfunctional states of aggregation-prone IDPs to be probed[52–54].

## smFRET experimental setup and protein labeling

smFRET experiments typically employ either Total Internal Reflection Fluorescence (TIRF) microscopy to observe immobilized molecules, or confocal spectroscopy to probe both freely diffusing and immobilized molecules. For IDPs, confocal-based smFRET offers several advantages, including the avoidance of perturbations from surface immobilization and the capability for multiparameter fluorescence detection, including wavelength, fluorescence lifetime, anisotropy, and correlation analyses[55].

Proteins are labeled site-specifically with donor-acceptor fluorophore pairs via cysteine residues or more advanced labeling strategies, such as non-natural amino acids or intein-based ligation methods[56,57]. Heterogeneously labeled populations and different stoichiometries of intermolecular FRET can be resolved by alternating laser excitation (ALEX) or pulsed interleaved excitation (PIE)[45,58]. The functional impact of labeling must be carefully assessed through control experiments, such as comparing the labeled and unlabeled IDP behavior in functional assays. Fluorophore mobility can be evaluated through time-resolved fluorescence anisotropy to ensure rotational freedom[44,45,55].

## Quantitative analysis of smFRET data for IDPs

The FRET efficiency, *E*, for IDPs cannot be interpreted in terms of a fixed donor-acceptor distance. Instead, IDPs require analytical approaches that can account for the broad distributions, $P(r)$, of the inter-fluorophore distances $r$, and the dynamic time scales of protein and dye motions. Suitable models from polymer physics have commonly been employed for describing global IDP features and to approximate $P(r)$[46]. Fluorescence lifetime measurements complement smFRET for accurate $P(r)$ determination. Two-dimensional plots of donor and acceptor fluorescence lifetime versus *E* can reveal whether a linear relationship exists (indicating fixed distances) or if deviations suggest broad and rapidly sampled distributions, offering a more stringent test of polymer models[44,45]. The integration with advanced computational approaches, including molecular simulations and Bayesian reweighting, can provide more detailed descriptions of $P(r)$ and deviations from simple polymer models, for instance in the presence of folded parts[47,59]. Nanosecond fluorescence correlation spectroscopy (nsFCS), combined with smFRET, enables the measurement of fast dynamics, including the reconfiguration times of disordered chains, which can be directly compared to simulations[46].

Important aspects for smFRET are to assess potential perturbations of the protein by fluorescent labeling, the analysis of complex distance distributions, and the role of the relevant timescales of the measurement for signal averaging[46,60]. Advances in labeling technologies and integration with computational modeling, including an explicit representation of fluorophores, and complementary biophysical techniques make smFRET invaluable in integrative approaches of IDP ensemble determination[61,62].

# Small-angle scattering (SAS) of X-rays and neutrons

Small-angle scattering (SAS) of X-rays (SAXS) and neutrons (SANS) have emerged as powerful techniques for the structural characterization of biomolecular systems in solution[63]. Despite being a low resolution (1-2 nm) technique, SAS offers valuable structural information on the overall size and shape of biomolecules and biomolecular assemblies. In the context of IDPs, this information is key to constraint the ensemble description and it is complementary to residue-level information derived from other techniques[64–66]. Major advances in beamline instrumentation, automatization, sample preparation and computational methods in the last decade have led to a tremendous increase in the application of SAS to structural biology[67]. The fast readout capabilities of current detectors enable kinetic monitoring of reactions across a broad time scale, from μs to hours, triggered by stopped-flow[68] or temperature jumps[69].

Although the conceptual framework of X-ray and neutron scattering as well as the strategies applied for their analysis are similar, these techniques differ in practical aspects, especially related to their sensitivity, which is much higher for SAXS[70,71]. The advantage of SANS arises from the different scattering power of nuclei, particularly large between the proton and the deuteron. Exploiting this isotope effect through contrast variation experiments of partially deuterated samples has been a key concept to enrich the structural content of SANS experiments[70,71].

## Analysis of SAS data

The scattering profile of an IDP is the average of those arising from all coexisting conformations sampled in solution. As a consequence, SAS profiles measured for IDPs do not display features along the momentum transfer ($q$) range[64]. Traditionally, Kratky plots ($I(q)\cdot q^2$ as a function of $q$), especially in their dimensionless form[65], have been used to qualitatively identify (partly) disordered states and distinguish them from globular particles. The Kratky representation has the capacity to enhance particular features of scattering profiles, allowing an easier identification of different degrees of compactness[72].

The radius of gyration, $R_g$, which can be obtained from the smallest angles of a SAS curve using the Guinier approximation, is the most common descriptor to quantify the overall size of particles in solution. The experimental $R_g$ is a single value representation of the size of the molecule, which for disordered states represents an ensemble average over all accessible conformations of the IDP and its hydration layer. The most common quantitative interpretation of $R_g$ for IDPs is based on Flory's equation, which relates it to the length of the protein chain through a power law ($R_g \propto N^{\upsilon}$), where $N$ is the number of residues and $\upsilon$ is an exponential scaling factor. For chemically denatured proteins a $\upsilon$ value of $\simeq$0.6 is expected[73]. However, according to theoretical models, IDPs are slightly more compact, with a $\upsilon$ value of 0.522 ± 0.01 [74]. Recently, an approach has been developed to simultaneously derive the $R_g$ and the $\upsilon$ value from a single SAXS profile[75], which has been used to connect the compactness of IDPs with their capacity to phase separate[36,76].

The pairwise distance distribution, *P(r)*, derived from the Fourier transformation of the SAS profile, provides a 1D representation of the molecular structure, and it can also be used to qualitatively identify protein flexibility. The *P(r)* also provides the maximum intramolecular distance, $D_{max}$, among all coexisting conformers. However, in the context of disordered proteins, $D_{max}$ is not a robust parameter and its quantitative interpretation is not recommended[77,78].

For IDPs, ensemble approaches are required, relying on two key ingredients: *(i)* accurate models of disordered states, and *(ii)* forward models with the capacity to robustly predict scattering properties including contributions from the solvent to the measured SAS data[79].

# Other biophysical techniques

Other complementary experimental methods have also gained attention in the studying of IDP structural properties (**Supplementary Table 1**). **Hydrogen deuterium exchange (HDX)** relies on measuring the rate at which amide protons in the protein backbone exchange with deuterium in a heavy water solution, a process highly sensitive to structural fluctuations and hydrogen bonding. For IDPs, HDX (HDX-NMR) by NMR spectroscopy or in particular by mass spectrometry (HDX-MS) provides valuable insights into disorder-to-order transitions, transient secondary structure formation, and conformational flexibility, making it a complementary technique to other structural methods like NMR, smFRET, or SAS. However, given the high exchange rates of IDPs, careful control of experimental conditions, such as pH, temperature, and labeling times, becomes critical to ensure meaningful interpretation of the data[80].

Mass spectrometry-based techniques beyond HDX-MS, such as hydroxyl radical footprinting (e.g. fast photochemical oxidation of proteins), diethylpyrocarbonate (DEPC) labeling, and carboxyl or lysine-specific covalent modifications, provide complementary insights into IDP dynamics by irreversibly labeling solvent-exposed side chains on rapid timescales[81]. These methods capture transient and heterogeneous conformations that HDX may miss, offering residue-level resolution of side-chain exposure. Applications include detecting structural transitions during amyloid formation[82], mapping interaction interfaces[83], and identifying microenvironment-sensitive regions[84,85]. When combined with modeling, these approaches enhance our ability to characterize the dynamic ensembles and functional states of IDPs.

**Circular dichroism (CD) spectroscopy** is a common technique for characterizing protein secondary structures by measuring the differential absorption of left- and right-circularly polarized light. Traditionally calibrated for globular proteins, CD spectroscopy has faced challenges in analyzing IDPs due to their weak and featureless spectra. Recent advancements have introduced new IDP reference datasets[86,87], enhancing the ability of this technique to detect conformational changes under various conditions. Despite these improvements, precise secondary structure quantification of IDPs still remains a challenge.

**Electron paramagnetic resonance (EPR) spectroscopy**, combined with site-directed spin labeling (SDSL), is used to detect unpaired electrons and measure distance distributions between specific residues. It has emerged as a powerful technique for studying IDPs[88]. SDSL-EPR is particularly useful for monitoring folding events and structural behavior, as it is not limited by protein size or system complexity[89]. The technique has been successfully applied to study disease-associated IDPs such as α-synuclein and tau, providing valuable insights into their conformational ensembles and interactions with the environment[88,90].

## Integrative experimental modeling

Integrative experimental approaches rely on combining complementary experimental data to determine conformational ensembles. An analysis of the Protein Ensemble Database (PED)[91] shows that the most common hybrid approaches combine SAXS with NMR and/or smFRET. Notably, Gomes et al. [92] used experimental NMR and SAXS data to construct the conformational ensembles for Sic1 and phosphorylated Sic1, and subsequently employed smFRET data as an independent validation criterion. In another recent study, Borthakur et al. [93] demonstrated that reweighted ensembles obtained from MD simulations using different force fields converge to highly similar ensembles when incorporating extensive experimental data from NMR and SAXS. In addition, other combinations can be useful, such as SAXS, smFRET, fluorescence correlation spectroscopy and dynamic light scattering, as shown by Borgia et al[94] . Such integrative studies illustrate how the information from different techniques is complementary.

## Bridging Experiments and Theory: Forward models

A critical element in translating experimental data into conformational ensembles is to define procedures, referred to as forward models, to calculate experimental observables from protein structures[95,96]. A major problem of forward models for IDPs is that most experimental methods provide data describing the average properties of proteins in solution, often over different timescales, rather than features of individual protein molecules, thus requiring a corresponding conformational averaging in the calculations[95,96]. Forward models estimating the experimental observables from generated models enable fitting experimental data obtained from various sources with a representative collection of conformations, as outlined in Table 1. Since the use of forward models influences the fitting procedure, selecting the appropriate model requires detailed understanding and considerations[97]. For instance, parameters of forward models need to be adjusted, ideally through self-consistent refinement against experimental datasets of both forward models and conformational ensembles[97–99]. Ultimately, experimental observations are influenced by both systematic and random errors, which must be accounted for to produce accurate and precise models[96].

To consider errors and the inherent variability of experimental observables, different Bayesian inference frameworks have been suggested[96,97]. Nonetheless, the development of more accurate and computationally efficient forward models and ML tools to account for different sources of error remains an active area of research.

**Table 1. Main experimental methods to determine structural ensembles of IDPs**

| Technique | Experimental measurement | Forward Models / Computational tools / Theoretical models | Information Content | Structural interpretation (e.g. residue level vs. global) | System perturbation | Environment (In-vivo / in-vitro / in-cell) | Reference database |
|---|---|---|---|---|---|---|---|
| NMR | Chemical shifts (traditional, 13C-NMR, …) | ShiftX, SHIFTS, Sparta, PPM1, CamShift | Local secondary structure propensity; ensemble-averaged deviations from random coil values. Limited | Dihedral angles, secondary structure population, dynamics | Isotope labelling | In-cell / in-vitro | bmrb.io |
| | | SSP, RCI, d2D, Shiftcrypt | | | | | |
| | PRE, PRI | dipole-dipole equation+correlation function, DEER-PREdict, MMM | Ensemble-averaged distances between spin labels and nuclei; provides insight into transient, weak, or low-population interactions and structural heterogeneity. | Distances/regions with correlated folding/dynamics | Chemical attachment of a paramagnetic label | | |
| | | covariance/correlation analysis | | | | | |
| | Relaxation rates | Model to data: molecular dynamics simulation+correlation function analysis+weighting | Time constants for intramolecular and global motions; order parameters reflecting flexibility; reveals timescales of molecular motions. | Correlation times+weights/order parameters | Isotope labelling | | |
| | | Data to model: Spectral density mapping, model-free analysis, IMPACT | | | | | |
| | DOSY | Data to model: Stokes-Einstein equation Model to data: Kirkwood-Riseman equation, HYDROPRO | Global hydrodynamic radius; informs on chain expansion or compaction | Translation diffusion coefficient | | | |
| | J couplings | Karplus equation | Backbone dihedral angles ($\phi$); detects transient secondary structure. | Dihedral angles | | | |
| | Cross-correlated | Data to Model: Correlation functions and | Dihedral angles based on correlated relaxation | Dihedral angles | | | |
| | cross-relaxation | Karplus-like equation | processes between two couplings. | | | | |
| | RDCs | Modeling alignment and calculation of the coupling | Bond vector orientation; indicates transient structural elements and long-range order | Mutual orientation | External alignment (mechanically stressed gels, alignment media) | | |
| SAXS | Scattering intensity as a function of the angle | CRYSOL/WAXSiS Pepsi-SAXS/FoXS | Provides global structural information | Global shapes and sizes, low resolution structures | None required | in-vitro | sasbdb.org |
| SANS | Scattering intensity as a function of the angle | CRYSON/pepsi-SANS | Complementary to SAXS | Global shapes and sizes, low resolution structures | Deuteration | in-vitro | sasbdb.org |
| (sm)FRET | Transfer efficiency (fluorescence bursts or trajectories) | Förster equation + polymer models, molecular simulations + dye model (e.g. FRET Positioning and Screening, Seidel), rotamer-library approaches (e.g. FRETpredict) | Number of populations, FRET efficiencies, fluorescence lifetimes, fluorescence anisotropies | Intra- or intermolecular distance/distance distribution, conformational changes | Dye labeling | In-vitro/in-cell | N/A |
| | Fluorescence lifetime decay | | | Intra- or intermolecular distance distribution | | | |
| | Time correlation (nanosecond FCS) | | | Distance relaxation time | | | |

# Computational generation of conformational ensembles

Generating conformational ensembles through computational techniques is a critical step in integrative modeling of IDPs, providing atomic-level insight that complements experimental measurements. Recent progress has been driven by tighter integration between experimental and modeling efforts, where experimental data refine models and models aid in the interpretation of complex experiments[95]. Three main strategies are commonly used to generate conformational ensembles: **knowledge-based methods**, which draw on empirical patterns or structural fragments; **physics-based simulations**, which are based on biophysical principles to sample the conformational landscape; and **machine learning–driven approaches**, which infer structural features from data without explicit physical modeling. Each strategy offers distinct advantages, and their selection depends on the level of detail required, available data, and modeling goals.

## Knowledge-based methods

Knowledge-based methods **(Fig. 2a)** generate conformational ensembles using heuristics or statistical patterns derived from known protein structures. These approaches often rely on fragment libraries, rotamer statistics, or backbone angle distributions informed by structural databases. Examples of such methods include Flexible-Meccano[100], TraDES[101], and IDPConformerGenerator[102], which employ fragments from non-redundant structural databases and statistical coil models to generate ensembles. Because they do not require energy calculations, they are computationally efficient and suitable for generating large conformational pools. These conformational collections can then be refined or reweighted based on experimental constraints to approximate statistical ensembles. Importantly, knowledge-based approaches are particularly useful when only limited experimental data are available, or when rapid conformational collection generation is needed for large-scale applications. A summary of representative methods and typical use cases is provided in **Supplementary Table 2**.

## Physics-based methods

Physics-based approaches **(Fig. 2b)** use molecular simulations to generate IDP ensembles. Since force fields originally developed for folded proteins often underestimate the flexibility and expanded conformations typical of disordered chains, IDP-specific force fields such as a99SB-disp[103] and Amber ff99SBws/ff03ws[104] have been developed to improve the balance between intramolecular and solvent interactions to better reflect IDP conformational preferences. Another challenge is sampling efficiency, as IDPs explore vast and shallow free energy landscapes. Techniques like replica exchange molecular dynamics (REMD)[105], which run simulations in parallel at different temperatures to enhance conformational exploration, are commonly used to improve sampling. More recently, hybrid strategies that combine simulation with machine learning (ML)—for example, using ML-derived potentials to guide sampling—have emerged as promising tools [106,107]. Alternative frameworks such as hydrophobicity scale (HPS) models[108] and CALVADOS [109,110], coarse-grained models parameterized against experimental data on IDPs and multidomain proteins, reduce computational cost while preserving some of the essential sequence-dependent behaviors. Beyond ensemble generation, physics-based methods increasingly contribute to design-oriented frameworks—from modeling the thermodynamic basis of multiphase condensates[111], to uncovering sequence–ensemble–function relationships in live cells[112] , and enabling inverse design of IDRs with target ensemble properties[113–115]. Representative methods and applications are summarized in **Supplementary Table 2**.

## Machine learning (ML) methods

ML approaches **(Fig. 2c)** have recently emerged as an alternative strategy for generating conformational ensembles of IDPs. These methods aim to learn sequence–ensemble relationships directly from structural data, providing a way to overcome some of the limitations of traditional simulation-based techniques. For example, idpGAN[116] is a generative adversarial network (GAN) trained on ensembles generated by CG molecular dynamics simulations, capable of producing novel ensembles with realistic $R_g$. While still in early stages, ML-based models offer a promising route for efficient ensemble generation, especially when integrated with simulation-derived[117,118] and/or experimental data[119]. In addition to generating ensembles, ML and related statistical parameter-learning methods can be used to tune force field parameters against experimental data[115,120], predicting IDP properties directly from sequence [4,6], designing novel IDP variants with desired properties[113,121] and predicting phase separation propensities of IDPs[122]. Representative ML-based methods for computational ensemble generation are summarized in **Supplementary Table 2**.

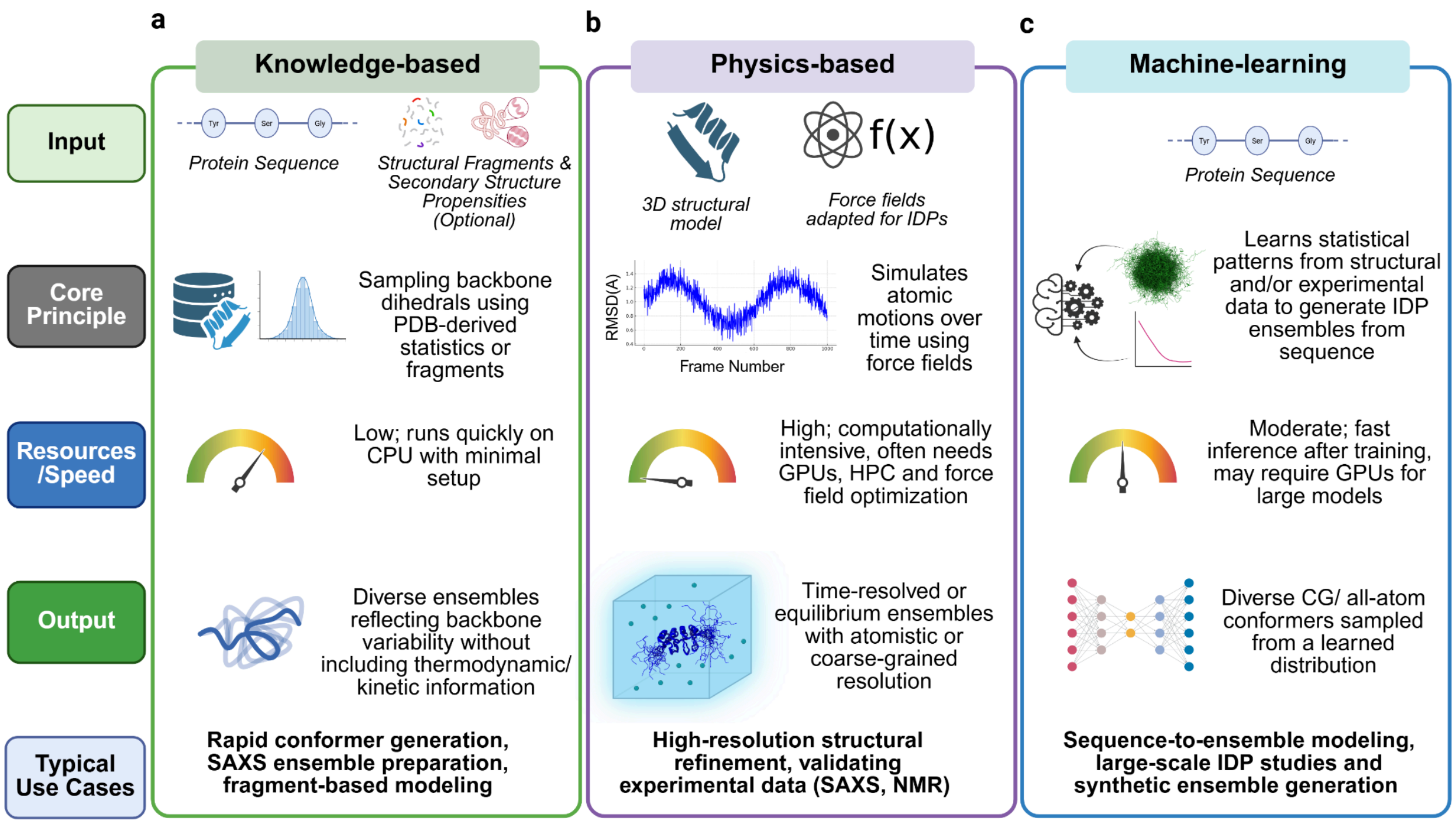

**Fig. 2. Schematic comparison of computational strategies for the generation of conformational ensembles of IDPs.** The diagram contrasts three broad methodological classes based on their inputs, underlying principles, computational demands, and typical applications. These approaches span from a) knowledge-based methods and b) physics-based simulations to modern c) machine learning techniques, each offering unique strengths for capturing IDP structural heterogeneity.

# Integration of computational models with experimental data

Experimental techniques such as NMR, SAXS, and smFRET provide important insights into the behavior of IDPs, but these measurements typically report on averages over a wide range of rapidly fluctuating conformations. By contrast, computational methods—such as molecular simulations—can generate detailed conformational ensembles from a protein sequence. However, discrepancies often arise between simulations and experiments, reflecting inaccuracies in force fields, limited sampling, or the nature of the experimental observable itself [123].

To address this aspect, an integrative modeling strategy is needed—one in which experimental data and computational ensembles inform and refine each other. The basic workflow **(Fig. 3)** involves generating an ensemble using a physics-based, knowledge-based or machine learning model, computing experimental observables from that ensemble using forward models, and comparing them with actual experimental measurements. When agreement is poor, the ensemble can be refined or reweighted so that its average properties better match the data.

Two main classes of approaches are available for this purpose. The first is based on maximum entropy (ME)[124], a foundational concept in statistical mechanics for generating the most unbiased ensemble consistent with known constraints. In its simplest form, and in the absence of experimental data, structural ensembles follow the Boltzmann distribution: $P_i = \frac{1}{Z}e^{-\beta E_i}$ where $E_i$ is the energy of conformation $i$ , $\beta$ is the inverse thermal energy, and $Z$ is the partition function. However, when experimental observables are introduced, such as NMR chemical shifts or SAXS profiles, this framework is generalized to a Gibbs distribution: $P_i = \frac{1}{Z}e^{-\Sigma_k \beta_k E_{ik}}$. Here, $E_{ik}$ represents an effective energy term for the deviation of conformation $i$ from experimental observable $k$, and $\beta_k$ (a Lagrange multiplier) adjusts the weight of that constraint. This formulation ensures the structural ensemble remains as close as possible to the original distribution while still fulfilling the experimental data. In practice, this statistical foundation, together with the principles from Bayesian statistics, underlies a number of ME-based methods such as Bayesian/Maximum Entropy (BME)[125], BioEn[126], and metainference [96]. These approaches implement the principle either by reweighting existing conformers post hoc, or by applying biasing forces during simulations to guide sampling. Such techniques are especially effective for IDPs, whose native states are characterized by extensive structural heterogeneity that must be preserved during model refinement.

The second category, known as maximum parsimony (MP) methods, seeks to identify the smallest subset of conformers from a large pool that can adequately explain the experimental data, minimizing structural complexity. In contrast to ME, MP methods aim to simplify the conformational collection by selecting or reweighting structures to maximize agreement with experimental observations while minimizing the size of the conformational collection. These approaches, such as EOM[127] and ASTEROIDS[128], are particularly useful when only a few conformational states are dominant and are thus most suitable for low-entropy systems with limited conformational heterogeneity, where a description in terms of statistical ensemble is not required.

Altogether, these integrative strategies allow researchers to model structural ensembles that are both physically realistic and experimentally validated, bridging the gap between simulation and experiments in the study of disordered proteins. Representative techniques for integration of experimental data are provided in **Supplementary Table 3**.

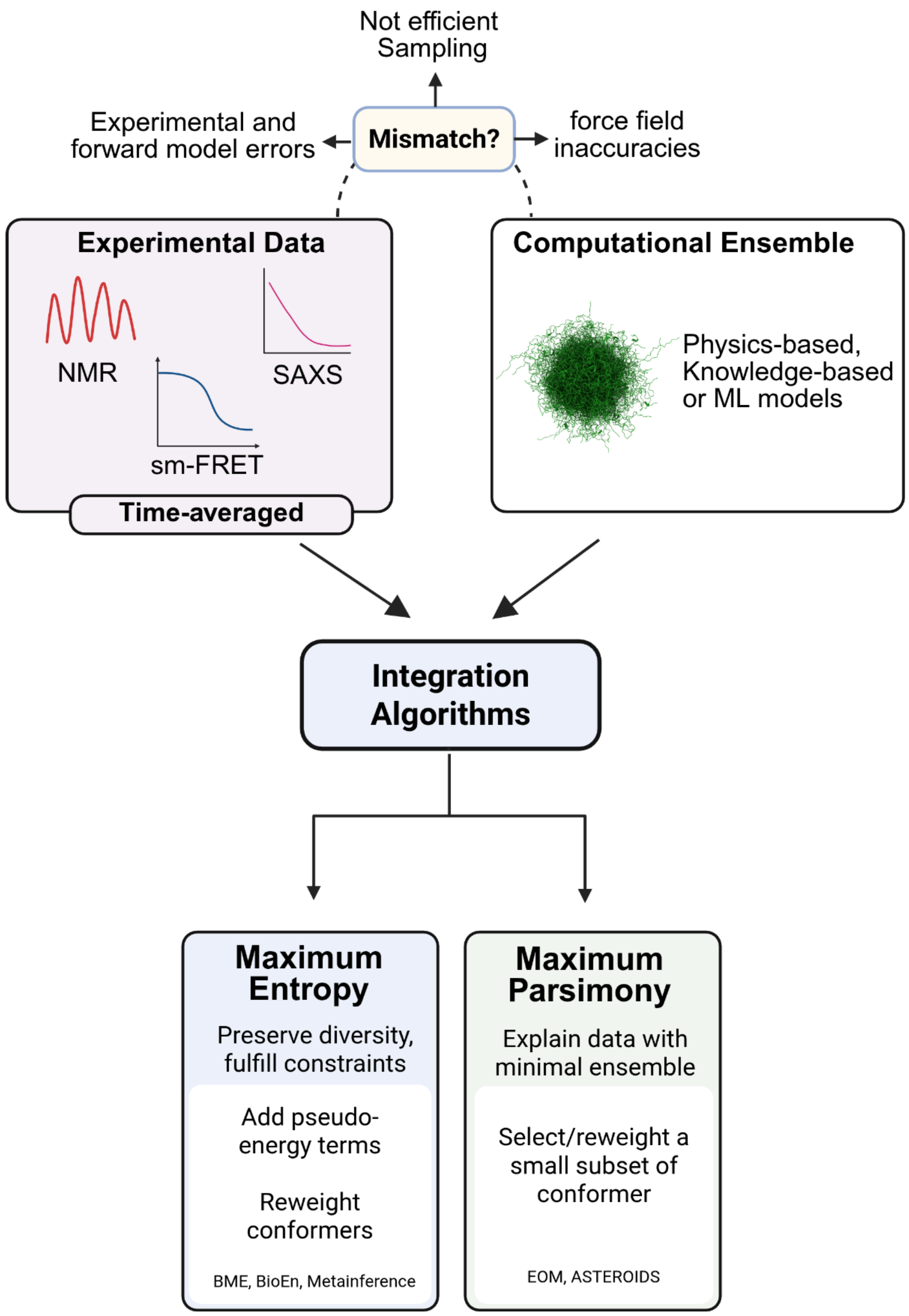

**Fig. 3**. **Overview of integrative modeling strategies for intrinsically disordered proteins (IDPs).** Experimental methods like NMR, SAXS, and smFRET provide averaged data across many conformations, while computational models generate detailed structural ensembles. When the two disagree—due to limited sampling, force field inaccuracies, or experimental noise—different strategies can be used to improve agreement. These include maximum entropy (ME) methods that adjust structural ensemble weights, maximum parsimony (MP) methods that generate conformational collections by selecting a minimal subset of structures. These strategies help align computational models with experimental data.

## Hybrid Computational techniques

Some computational methods for modeling IDPs leverage complementary modeling strategies. These hybrid approaches (**Supplementary Table 2**) integrate complementary ensemble generation techniques, or couple ensemble generation directly with experimental data integration strategies. One strategy involves combining physics-based and knowledge-based approaches to enhance sampling efficiency and structural diversity. For example, FastFloppyTail[129] employs fragment-based backbone sampling alongside energy-based refinement. Other techniques couple machine learning with physics-based modeling, such as AlphaFold–Metainference[130] and bAIes[107] which refines deep learning predictions from AlphaFold via Bayesian ensemble inference and molecular simulation. A different class of hybrid methods links ensemble generation with experimentally data-driven refinement. For instance, DynamICE [119] integrates machine learning with maximum entropy principles to iteratively optimize ensemble structures based on NMR data. These flexible and composable pipelines offer practical advantages when addressing the complexity of IDP ensembles, particularly in cases where no single modeling paradigm is sufficient.

## Open challenges in conformational ensemble generation

Despite recent advances, several challenges remain in generating conformational ensembles[123]. First, force field accuracy continues to be a critical limitation. While parametrizations using experimental data and quantum chemistry calculations are continuously improving their accuracy, current force fields still rely on rather simple functional forms established decades ago, and fail to describe subtle interactions, such as those arising from polarization effects, especially in complex or highly flexible biomolecules. Progress in developing more accurate force fields may require closer collaboration between experimental and computational researchers and the development of top-down approaches, including the adoption of machine-learning methods[109,120,131].

Second, the optimal integration of experimental data into simulations is still an open problem[123],[96]. Many forward models used to translate experimental data into simulation restraints are limited by their assumptions and approximations. Accounting for the ensemble-averaged nature of most experimental observables on IDPs, such as those derived from NMR or smFRET, or the effect of the hydration for SAS, is critical for accurate modeling. Robust frameworks that address uncertainties in both experimental data and forward models are essential to bridge this gap[96,123]. In the absence of ground-truth conformational ensembles, such frameworks require experimental data from orthogonal techniques and for different proteins to self-consistently select, refine, and validate forward models[97,99,132]. These experimental data are typically insufficient to describe a unique conformational ensemble, with the final result dependent on the computational model used to generate the ensemble.

Third, the efficient sampling of conformational space also remains a significant bottleneck. Insufficient sampling can lead to incomplete or biased conclusions, making it crucial to assess convergence rigorously. Advancements in some techniques, such as accelerated molecular dynamics and Markov state models[133,134], offer potential solutions but require widespread adoption and careful validation.

# Conformational ensemble validation and comparison

## Comparison of conformational ensembles

The criteria and metrics for validating and comparing structures of well-folded proteins determined by X-ray crystallography, NMR spectroscopy, cryo-EM/ET, integrative approaches, or predicted by ML models, like AlphaFold[135], are well established. Validation checks include evaluating bond angles, torsion angles, and ensuring that the protein adheres to known stereochemical properties. Tools like MolProbity[136] are often used for these validations. For comparison, the most widespread approach is based on structural alignment and quantitative RMSD (Root Mean Square Deviation) analysis[137]. Other scores to quantify the differences between protein structures (usually between predictions and experimental structures), such as the GDT (Global Distance Test)[138] or the TM-score (Template Modeling Score)[139], are frequently used, in particular to assess predictions in CASP (Critical Assessment of Structure Prediction)[140].

In contrast, the structural analysis of conformational ensembles of IDPs is more complex. As most experimental methods (Table 1) provide averaged values of measurements, these are hard to connect to usual structural descriptors, and in particular to atomic coordinates. This represents a great challenge for the evaluation of computational methods that generate conformational ensembles as a weighted set of conformers relying on atomistic or coarse-grained models[95,141], as those deposited in the Protein Ensemble Database (PED)[91].

The comparison of conformational ensembles, however, serves multiple purposes. The determination of an IDP ensemble from sparse experimental data is a mathematically underdetermined problem, leading to multiple solutions. The comparison of these ensembles may be critical for validating them (see next section), and/or for deriving mechanistic insights by identifying functional correlates (e.g. binding regions, dynamic linkers), comparing homologous IDPs and assessing the influence of solution conditions. As conformational ensemble comparison methods for folded proteins[142] are difficult to generalize to IDPs, various approaches have been developed in recent years. A conformational ensemble can be represented by a single scalar that quantifies its conformational heterogeneity[143], and differences in this value can be used to compare two conformational ensembles. While this metric is easy to interpret, it offers only a limited view of the differences between conformational ensembles and lacks the resolution to capture more detailed structural or dynamical variations. A more sophisticated approach involves analyzing the differences in distance distributions of individual residue pairs and assessing their statistical significance between two conformational ensembles[144]. However, given the vast conformational variability of IDPs, it is essential to consider both average properties and their distributions. Indeed, reducing IDP conformational descriptors to their mean can lead to significant information loss. Several methods have been proposed for extending comparative structural analyses to a distributional framework, using various similarity measures to compare probability distribution functions, such as the Kullback-Leibler and the Jensen-Shannon divergence[142,144,145], Hellinger distance[146] or the Wasserstein distance[147]. Among these, the Wasserstein distance is particularly appealing for its incorporation of the geometry of the underlying space, though it is more computationally demanding than the others. In this framework, different descriptors can be used to characterize conformations inside each ensemble and to compute pairwise distances. The most natural one is to use atomic coordinates and RMSD[142], however, the required structural superposition hinders its application to IDPs. To circumvent this problem, other descriptors such as torsion angles[148] or other local structural properties[145] can be considered. For a more comprehensive comparison, local and global structural descriptors can be taken into account in a unified way[147]. Other approaches for the comparative analysis of conformational ensembles, such as methods incorporating topology concepts[149] or the identification of local energy traps at the residue level[150], have been explored. **Supplementary Table 4** summarizes conformational ensemble comparison methods.

## Validation of conformational ensembles

As discussed in the previous sections, there are several ways — experimental, computational, or integrated — of determining conformational ensembles of IDPs. The comparison of alternative solutions is not trivial, and it also impinges upon our ability to select for the correct ones that reflect all important structural aspects of the actual conformational ensembles. Although it is very challenging to know the accuracy and precision of a conformational ensemble, objective criteria to assess its quality can be defined. In general, a conformational ensemble can be validated by experimental datausing orthogonal experimental techniques, pointing toward the use of integrative approaches.

The general path to validation begins with ascertaining that the IDP ensemble adheres to simple physicochemical and stereochemical rules, and it complies with all ensemble-related experimental data. That is, the first step is to check if all conformations in the ensemble comply with measures of bond lengths and torsion angles, location on the Ramachandran map, and avoidance of atomic clashes. These are all conditions not always currently met in PED ensembles[91].

The next step is to check the adherence of the ensemble to mean global structural parameters derived from experimental techniques, primarily SAS and smFRET, or calculated from sequence by applying polymer-physical principles[4,6,46], such as the $R_g$, end-to-end distance, geometric deviation from a perfect sphere (asphericity) and scaling exponents ($\upsilon$)[151–154]. As such global parameters may be compromised when secondary structures or long-range contacts are present, one should proceed from global descriptors to more local conformational parameters, such as short- and medium-range distance information (coming from smFRET and PRE), descriptors based on inter-residue contact probabilities[155,156], secondary-structure content (as reported by CD or NMR), and local residue-specific conformational preferences (Ramachandran torsion angles φ and ψ), which can be derived primarily from NMR data (CSs, RDCs, NOEs).

The next step is to check for the robustness of a conformational ensemble, by assessing how it depends on particular data points. A sign of limited accuracy is large alterations of the ensemble upon removing part of the input data[123,157]. On the more positive side, the quality of the conformational ensemble can be ascertained if the ensemble can predict independent experimental observables[158], as demonstrated in two different settings. In one study, representative ensembles of α-synuclein (140 amino acids) and tau protein (441 amino acids) were derived from NMR and SAXS data. These conformational ensembles were cross-validated by selectively omitting portions of the data during the ensemble selection procedure and subsequently back-calculating the omitted data from the resulting ensembles[159]. This approach was shown to lead to more accurate predictions than those from a statistical coil model. In another similar experiment, the structural ensemble of the disordered N-terminus of measles virus phosphoprotein (110 amino acids) was approached by a variety of NMR, SAXS and smFRET data, with particular emphasis on the interplay of NMR paramagnetic relaxation enhancement (PRE) and smFRET data expectedly providing similar constraints on medium- and long-range distance distributions[128]. Interestingly, the conformational ensembles calculated by leaving one type of data out – PRE or smFRET – had some power in reproducing the other, but the full complement of data was required for reproducing all aspects of the ensemble.

A more abstract structural feature to judge for validation is if the conformational ensemble has some biological/functional relevance. If an IDP interacts with other molecules (e.g., proteins, DNA, RNA), one may assess its ability to recapitulate the experimentally observed binding interfaces and binding modes. A particular feature toward this goal is that IDPs often function by molecular recognition undergoing disorder to order transition upon binding which can assume an unbound structural state similar to their partner bound conformation[160]. Capturing this feature by the conformational ensemble is a sign of its functional relevance[144,161]. Recently, AlphaFold has been shown to reliably predict relevant binding modes of disordered regions[162]. There are also cases where the global dimensions of linker regions have been shown to be evolutionarily conserved[4,5]. Additionally, correlations have been observed between IDP compaction and gene-ontology (GO) protein function descriptors [4].

In a similar vein, one might check for disease relevance. If an IDP is involved in disease, one may consider if its conformational ensemble provides insights into disease mechanisms and/or potential therapeutic targeting. To this end, one may ask if *in silico* docked small molecules do experimentally bind, whereas extending these to broader structure-activity-relationships (SAR) of molecular analogues has even more confirmatory power[163–165].

As suggested above, determining conformational ensembles from sparse experimental data is a mathematically ill-posed problem, which has multiple solutions[123]. However, from an experimental point of view, the problem becomes well-posed if different conformational ensembles are experimentally indistinguishable[4,123].

Looking forward, we note that the conformational ensembles that we aim to compare and validate are determined mostly under *in vitro* conditions, and thus they may not adequately reflect the true structural preferences of the proteins in their physiological, cellular milieu. That is, the ultimate verification of a conformational ensemble is to validate it against cellular structural data, for example obtained by in-cell NMR or smFRET[51,166], which are still in their infancy. It has already been shown, however, that the structural disorder of several IDPs persists in mammalian cells and show behavior close to that recorded *in vitro*, e.g. α-synuclein investigated by NMR[41], prothymosin α observed with smFRET[51],or comparisons between in vitro and in vivo FRET measurements[167]. Such approaches should be routinely invoked as a conclusive element of IDP ensemble validation. Another approach involves using analyses of sequence conservation signals to gauge the conformational properties of IDPs that are important for function in living organisms[168–170] .

# Roadmap for conformational ensemble generation

The determination of conformational ensembles of IDPs stands at a critical juncture, where methodological diversity is expanding rapidly but standardized practices for validation, comparison, and application remain elusive. To address these challenges, we propose a four-pronged roadmap aimed at shaping a rigorous and accessible future for IDP ensemble modeling. This includes: (1) establishing a benchmarking infrastructure to drive community-wide progress, (2) developing robust methods for quantifying and propagating uncertainty in ensemble generation and evaluation, (3) incorporating strategies for capturing non-equilibrium dynamics and kinetic features of disordered systems, and (4) offering practical guidelines for method selection.

## 1. IDP Ensemble Benchmarking Challenge (IDP-Bench)

A cornerstone for advancing the field is the creation of an open, transparent, and standardized benchmarking initiative. Inspired by the success of CASP[171] in protein structure prediction and of CAID in the critical assessment of intrinsic disorder prediction[172], we introduce the IDP Ensemble Benchmarking Challenge (IDP-Bench), a recurring, community-driven assessment platform specifically tailored to IDP structural ensembles. IDP-Bench will be structured as a four-phase biennial cycle aligned with CAID initiative (**Supplementary Fig. 1**: (i) selection and curation of high-quality, diverse experimental datasets by an independent committee, (ii) submission of ensemble predictions by participating groups, (iii) automated, multi-metric scoring against withheld validation data, and (iv) continuous public dissemination of results via a dedicated website. The first edition will evaluate predictions for well-characterized IDPs, such as α-synuclein, Sic1, the K18 repeat domain of tau and full-length tau, p27Kip1, and the measles-virus NTAIL domain, each with extensive public datasets (SAXS, PRE, RDC, smFRET, HDX-MS). These proteins collectively span different sequence features, biological functions, and levels of experimental accessibility, and none has yet been evaluated under a uniform blind protocol.

Crucially, scoring metrics will assess both global (e.g., $\chi^2$ for SAXS profiles, KL divergence for P(r)) and local agreement (e.g., PRE RMSE, RDC Q-factor, smFRET Wasserstein distance) (see **Supplementary Table 5)**. These will be combined into a weighted composite score, with full scoring scripts and error models released as open-source tools.

A blind component will be introduced further in IDP-Bench, e.g., SAXS or NMR data acquired under a given set of conditions, in order to test true predictive capability.

Altogether, this initiative lays the foundation for a sustainable benchmarking ecosystem—one that promotes transparency, drives methodological innovation, and enables long-term tracking of progress in IDP ensemble prediction.

## 2. Uncertainty in conformational ensemble determination

Benchmarking conformational ensembles of IDPs requires not only standardized data and scoring metrics, but also a rigorous framework for handling uncertainty. We therefore plan to integrate uncertainty quantification directly into the IDP-Bench framework, making it a foundational criterion for evaluating the quality of the conformational ensembles. **Supplementary Fig. 2** summarizes the error landscape in the determination of IDP ensembles, focusing on key experimental observables. It serves as a baseline guide for experimental reproducibility and model fitting. We also should consider forward-model error in our variance estimates. For instance, chemical shift back-calculations using SPARTA+[17] yield a standard deviation of about 1 ppm for backbone Cα predictions on its training dataset[173], and SAXS profile predictions evaluated against benchmark consensus data reveal that even widely used tools like CRYSOL [174] can show systematic deviations due to differences in hydration layer modeling and background correction[175]. Where empirical estimates are lacking, we advocate bootstrapping over modeling parameters (e.g., fluorophore grids, rotamer libraries) to obtain robust confidence intervals. The IDP-Bench scoring pipeline will explicitly output both standard and scoring metrics: reduced $\chi^2$, residual distributions, and Bayes factors. These help distinguish between genuine fits and artifacts of overfitting. To enforce transparency, future participants will submit a standardized reporting checklist detailing: (i) experimental data with point-wise uncertainties, (ii) forward-model error sources and propagation strategies, (iii) bootstrap/posterior distributions for fitted parameters, and (iv) treatment of outliers.
By embedding uncertainty quantification into every stage, from experimental input to scoring and reporting, we ensure that ensemble accuracy is assessed not only in absolute terms, but relative to the data's inherent noise and modeling limitations.

## 3. Dynamic properties and non-equilibrium behavior of IDPs

Non-equilibrium dynamical processes represent a promising but still nascent area in IDP structural biology. Many IDPs function through transient interactions and conformational switching, which are not adequately captured by equilibrium models. A major challenge is the acquisition of time-resolved experimental data, but recent advances in techniques such as time-resolved X-ray crystallography with free-electron lasers[176], real-time NMR[177], time-resolved and single-molecule FRET[178], time-resolved HDX-MS[179] and SAXS[180], and liquid-state electron microscopy[181] are beginning to enable such studies. These methods are increasingly being used to provide structural restraints for simulations that aim to reconstruct the time evolution of IDP ensembles.
To model these processes, recent approaches apply frameworks based on maximum entropy and its dynamical extension, maximum caliber. These allow for the inference of IDP dynamics consistent with both experimental data and physical principles, enabling the modeling of kinetic pathways and transient conformational states[182,183]. While still in early development, this direction offers a path toward dynamic ensemble models that more closely reflect the time-dependent behavior of IDPs in their functional contexts.

### 4. Guidelines for method selection

Beyond standardization and validation, progress in IDP ensemble determination requires clear guidance for choosing the appropriate tools and strategies based on available data, biological questions, and computational resources. To this end, we propose a pragmatic decision framework for method selection, derived from the typical reasoning process of experienced practitioners. The guidelines begin by evaluating available experimental data. When no structural data are available, molecular dynamics simulations (e.g., coarse-grained models or ML-generated backbones) can offer baseline hypotheses. For systems constrained only by local observables (e.g., chemical shifts), we recommend knowledge-based or ML-based backbone generators coupled with reweighting. When both local and global data (e.g., smFRET, SAXS) are available, hybrid restrained simulations or Bayesian/maximum-entropy methods offer optimal balance between fidelity and generalizability. Next, the biological goal should guide model resolution. All-atom models are ideal for site-specific predictions (e.g., binding interfaces, mutational effects), while coarse-grained models excel in large-scale problems (e.g., phase behavior, scaling laws) where speed and statistical representativeness are key.

Finally, an important consideration in method selection is the available computational resources. For projects with limited resources, such as a desktop workstation, methods based on accelerated Monte Carlo sampling or knowledge-based generation may be most appropriate. Access to a modest GPU cluster enables more advanced techniques, including replica-exchange molecular dynamics or hybrid reweighting approaches. For users with access to high-performance computing infrastructure, all-atom replica simulations and machine learning–based ensemble refinement methods become feasible options.

These decision rules are captured in a simplified flowchart **(Supplementary Fig. 3)** and two supplementary tables: (i) ensemble generation tools cross-referenced with their input types, compute cost, and best-use cases **(Supplementary Table 2),** and (ii) data integration tools indicating whether they propagate uncertainty and accommodate forward-model variability **(Supplementary Table 3).**

## Conclusion

In the face of growing interest in disordered proteins and the understating of their structural properties, this article establishes a comprehensive and practical framework to guide IDP ensemble determination, benchmarking, and interpretation. By integrating community expertise across experimental and computational domains, we chart a roadmap for IDP ensemble determination, emphasize the need for uncertainty quantification, and propose actionable benchmarking strategies. To implement this roadmap, we established the IDP Ensemble Benchmarking Challenge (IDP-Bench), following the examples of CASP and CAID, which greatly contributed to major advances in computational structural biology. This collaborative effort not only provides clarity on the current landscape of conformational ensemble modeling but also defines a shared vision for its future. As the field moves toward consensus and interoperability, we hope this work will serve as both a reference and a call to action, catalyzing rigorous, reproducible, and biologically relevant studies of disordered protein conformational space.

# Acknowledgments

This work has been supported by: COST Action ML4NGP [CA21160], supported by COST (European Cooperation in Science and Technology); European Union through NextGenerationEU; PNRR project ELIXIRxNextGenIT [IR0000010]; National Center for Gene Therapy and Drugs based on RNA Technology [CN00000041]; European Union [101160233] (HORIZON-Twinning project IDP2Biomed) and under grant agreement no. 101182949 (HORIZON-MSCA-SE project IDPfun2); Ministry of Education, Youth and Sport of the Czech Republic, within program Inter-Excellence II, INTER-COST (LUC23081); 2022.01454.PTDC (from FCT-Portugal); Maria de Sousa Award 45/2022 (Bial Foundation and PT Medical Association); French National Research Agency (ANR) grant [ANR-22-CE45-0003] and [ANR-24-CE11-2382-03]; NKFIH RGH_24 (RGH 151464); ItaliaDomani PNRR project 'Potentiating the Italian Capacity for Structural Biology Services in Instruct-ERIC' (ITACA.SB, no. IR0000009); NKFIH RGH_24 (RGH 151464).
Views and opinions expressed are however those of the author(s) only and do not necessarily reflect those of the European Union or the European Research Executive Agency. Neither the European Union nor the granting authority can be held responsible for them.

# Author contributions

All authors contributed to the discussion and the initial draft of this manuscript. H.G wrote the final document with the help of the co-authors. All authors edited and refined the final manuscript. A.M.M and S.C.E.T coordinated the project.

# Competing interests

K.L.-L. holds stock options in and is a consultant for Peptone Ltd. All other authors declare no competing interests.

# Additional information

Correspondence should be addressed to Alexander Miguel Monzon or Silvio C. E. Tosatto